\documentclass{aastex61}

\usepackage{CJK}
\usepackage{amssymb}
\usepackage{multirow}
\usepackage{morefloats}
\usepackage{amsmath}
\usepackage{bigstrut}
\usepackage{booktabs}
\usepackage{natbib}
\usepackage{float}
\usepackage{graphicx,epsfig,fancyhdr,epsf,txfonts,epstopdf}
\usepackage{latexsym,bbm}
\usepackage{lineno}
\usepackage{color}
\usepackage{ulem}
\usepackage{threeparttable}
\usepackage{longtable}
\usepackage{hyperref}

\usepackage{lineno}

\received{***}
\revised{***}
\accepted{***}

\shortauthors{Li et al.}
\shorttitle{High-harmonic Plasma Emission}
\begin{document}
\large
\begin{CJK*}{UTF8}{gbsn}

\title{High-harmonic Plasma Emission Induced by Electron Beams in Weakly Magnetized Plasmas}

\correspondingauthor{Yao Chen}
\email{yaochen@sdu.edu.cn}

\author{Chuanyang Li (李传洋)}
\affiliation{Institute of Frontier and Interdisciplinary Science, Shandong University, Qingdao, Shandong, 266237, People's Republic of China}
\affiliation{College of Physics and Electronic information, Dezhou University, Dezhou, Shandong, 253023, People's Republic of China}

\author{Yao Chen (陈耀)}
\affiliation{Institute of Frontier and Interdisciplinary Science, Shandong University, Qingdao, Shandong, 266237, People's Republic of China}
\affiliation{Institute of Space Sciences, Shandong University, Shandong, 264209, People's Republic of China}

\author{Zilong Zhang (张子龙)}
\affiliation{Institute of Space Sciences, Shandong University, Shandong, 264209, People's Republic of China}

\author{Hao Ning (宁昊)}
\affiliation{Institute of Frontier and Interdisciplinary Science, Shandong University, Qingdao, Shandong, 266237, People's Republic of China}

\author{TangMu Li (李汤姆)}
\affiliation{Institute of Frontier and Interdisciplinary Science, Shandong University, Qingdao, Shandong, 266237, People's Republic of China}


\begin{abstract}
\large

Electromagnetic radiation at higher harmonics of the plasma frequency ($\omega \sim n\omega_{pe}, n > 2$) has been occasionally observed in type II and type III solar radio bursts, yet the underlying mechanism remains undetermined. Here we present two-dimensional fully kinetic electromagnetic particle-in-cell simulations with high spectral resolution to investigate the beam-driven plasma emission process in weakly magnetized plasmas of typical coronal conditions. We focused on the generation mechanisms of high-harmonic emission. We found that a larger beam velocity ($u_d$) favors the generation of the higher-harmonic emission. The emissions grow later for higher harmonics and decrease in intensity by $\sim$2 orders of magnitude for each jump of the harmonic number. The second and third harmonic ($\rm H_2$ and $\rm H_3$) emissions get closer in intensity with larger $u_d$. We also show that (1) the $\rm H_3$ emission is mainly generated via the coalescence of the $\rm H_2$ emission with the Langmuir waves, i.e., $\rm H_2 + L \rightarrow H_3$, wherein the coalescence with the forward-propagating beam-Langmuir wave leads to the forward-propagating $\rm H_3$, and coalescence with the backward-propagating Langmuir wave leads to the backward-propagating $\rm H_3$; and (2) the $\rm H_4$ emission mainly arises from the coalescence of the $\rm H_3$ emission with the forward- (backward-) propagating Langmuir wave, in terms of $\rm H_3 + L \rightarrow H_4$.
\end{abstract}

\keywords{
\href{http://astrothesaurus.org/uat/1261}{Plasma astrophysics (1261)};
\href{http://astrothesaurus.org/uat/1490}{Solar electromagnetic emission(1490)};
\href{http://astrothesaurus.org/uat/1522}{Solar radio emission(1522)};
\href{http://astrothesaurus.org/uat/1339}{Radio bursts (1339)};
\href{http://astrothesaurus.org/uat/1544}{Space plasmas(1544)};
\href{http://astrothesaurus.org/uat/1993}{Solar coronal radio emission (1993)};
}

\section{Introduction}

Plasma emission (PE) is defined as electromagnetic (EM) radiation at frequencies close to the plasma frequency $\omega_{pe}$ (F) and its second harmonic ($\rm H_2$). Several types of solar radio burst, the radio radiation from the outer heliospheric boundary, and some radio emissions from the planetary magnetosphere have been understood as PE (e.g., \citealp{Etcheto1984, Cairns1995, Cairns1998, Gurnett1998, Cairns2002, Kuncic2005, Chen2014, Vasanth2016, Vasanth2019, Lv2017, Li2017, Pisa2017, Tasnim2022}). Emission, probably at the third harmonic ($\rm H_3$), has been occasionally reported for type II (\citealp{Bakunin1990, Kliem1992, Zlotnik1998, Brazhenko2012}) and type III solar radio bursts (\citealp{Kundu1965, Takakura1974, Zlotnik1978, Cairns1986, Reiner1992, Reiner2019}). For other examples, \cite{Cairns1986} reported harmonics at $n\omega_{pe}$ ($n$ = 3--5) in the foreshock region of the bow shock with the ISEE1 data, while \cite{Reiner2019} detected possible $\rm H_3$ emissions in interplanetary type III bursts based on the Wind data.

\cite{Ginzburg1958} proposed the original framework of the theory of PE, which has been developed into the standard PE model (e.g., \citealp{Melrose1970a,Melrose1970b, Zheleznyakov1970a,Zheleznyakov1970b,Melrose1980, Melrose1987, Cairns1987, Robinson1994}). The model involves a multistep nonlinear process of wave-particle and wave-wave interactions: (1) efficient excitation of Langmuir (L) turbulence by electron beams through the kinetic bump-on-tail instability, (2) scattering of L waves by ion-acoustic (IA) waves or ion density inhomogeneities to generate the fundamental emission and/or backward-propagating Langmuir ($\rm L^\prime$) waves ($\rm L \pm IA \rightarrow F$ and $\rm L \pm IA \rightarrow L^\prime$), and (3) resonant coupling of forward- and backward-propagating Langmuir turbulence to generate the harmonic emission ($\rm L + L^\prime \rightarrow H_2$). The standard PE theory has been widely used to explain radio bursts in space and astrophysical plasmas, such as solar radio bursts in terms of types I--V (see, e.g., \citealp{Chen2014, Vasanth2016, Lv2017, Li2017, Vasanth2019}). Numerical verifications based on fully kinetic EM particle-in-cell (PIC) simulations (e.g., \citealp{Thurgood2015, Che2017, Henri2019, Ni2020, Chen2022, Zhang2022}) or weak turbulence theory have been performed (e.g., \citealp{Yoon2000, Yoon2005, Yoon2006, Yoon2012, Ziebell2015, Ziebell2016, Lee2019, Lee2022}).

However, the generation mechanism of high-harmonic emissions ($\rm H_3, H_4, \cdots$) remains inconclusive (\citealp{Zheleznyakov1974, Cairns1987, Yin1998, Zlotnik1998, Ziebell2015}). Three different mechanisms have been proposed: (1) coalescence of three L waves in terms of $\rm L + L^{\prime} + L^{\prime \prime} \rightarrow H_3$ (\citealp{Kliem1992}); (2) coalescence of an L wave with $\rm H_2$ in terms of $\rm L + H_2 \rightarrow H_3$ (\citealp{Zlotnik1978}), where this mechanism has been generalized by \cite{Cairns1988} as $\rm L + H_n \rightarrow H_{n+1}$; and (3) $\rm L + L_n \rightarrow H_{n+1}$ (\citealp{Yi2007}), i.e., coalescence of the primary L wave with the $n$th-harmonic L mode ($\rm L_n$).

Earlier numerical studies mainly focused on the generation of the F and $\rm H_2$ emissions (e.g., \citealp{Pritchett1983, Yin1998, Kasaba2001}), with few reports on high-harmonic emission. Using the two-dimensional relativistic EM PIC code to simulate the beam-plasma interaction process (\citealp{Matsumoto1993}), \cite{Rhee2009} found that high-harmonic emission (up to the fifth) can be generated with the beam velocity exceeding $0.5c$. By examining the radiation patterns and the temporal correlation between the electrostatic (ES) and EM waves, they suggested that the $\rm H_3$ ($\rm H_4$) emission resulted from the merging of the $\rm H_2$ ($\rm H_3$) emission with the L waves, in agreement with the second process mentioned above. Note that \cite{Rhee2009} utilized a relatively small domain ($512 \ \lambda_{De} \ \times \ 512 \  \lambda_{De}$, where $\lambda_{De}$ is the electron Debye length) and a limited duration (328 $\omega_{pe}^{-1}$) with a relatively low number of macroparticles (the number of macroparticles per cell per species, NPPCPS, which, for background electrons and protons and beam electrons, was 80, 80, and 8, respectively). These limitations led to excessive numerical noise along dispersion curves and a limited resolution. Thus, further validation is required.

Recently, \cite{Krafft2022} conducted two-dimensional PIC simulations in weakly magnetized plasmas using larger-domain ($1448 \ \lambda_{De} \ \times \ 1448 \  \lambda_{De}$), longer-duration (9000 $\omega_{pe}^{-1}$), and larger NPPCPS values (1800, 1800, 1800). Their study focused on the coalescence processes to generate $\rm H_3$ and $\rm H_4$ in the beam-plasma system with or without density fluctuations. They suggested the following processes for $\rm H_3$: $\rm H_2 + L^{\prime} (L^{\prime \prime}) \rightarrow H_3$, where $\rm L^{\prime} (L^{\prime \prime})$ is supposed to be the product of the first (second) cascade of the ES decay of the beam-Langmuir (BL) waves ($\rm L \rightarrow L^{\prime} + S^{\prime}$ and $\rm L^{\prime} \rightarrow L^{\prime \prime} + S^{\prime \prime}$); a very similar process has been suggested for $\rm H_4$. They did not reject the possibility of other processes such as $\rm L_2^{\prime} + L \rightarrow H_3$ and $\rm L_3^{\prime} + L \rightarrow H_4$.

With careful evaluations, we suggest that the result of \cite{Krafft2022} remains questionable and should be further tested due to the following concerns.

(1) Regarding the Fourier analysis used to obtain the high-harmonic spectra in the wavevector space, the authors restricted the frequency range to a small window centered around $\omega_n \approx  n\omega_{pe}$ (refer to their Figures 1(c) and (d) and 5(c) and (d)). This means only signals within the prescribed narrow range have been processed. Such analysis could not tell the existence of high-harmonic radiation since, even in the thermal case, such analysis can give enhanced signals around $\omega_n$, or any frequency ($> \omega_{pe}$). This is due to the local enhancement of numerical noise along dispersion curves with PIC simulations. We have confirmed this with our result, following the same procedure as employed by \cite{Krafft2022}.

(2) Regarding the uncertainties of the energy evaluation of $\rm H_4$, the $\rm H_4$ intensity is only marginally larger than the initial value in their homogeneous case, with no discernable increase of $\rm H_4$ intensity in the other case with density fluctuations (see Figures 2(a) and 4(a) in \citealp{Krafft2022}). This raises doubts as to whether the obtained $\rm H_4$ emission is really excited by the beam-plasma interaction or is simply a part of the numerical noise.

Another point is that the $\rm H_3$ and $\rm H_4$ emissions can be excited at a beam velocity of $0.25c$ according to \cite{Krafft2022}. This contradicts the abovementioned result of \cite{Rhee2009}. To clarify these issues, we repeated the simulations of the beam interaction with a uniform plasma, with basically the same configurations as used by \cite{Krafft2022}, such as the simulation domain, grid spacing, number of particles, etc.. This is done with the Vector-PIC (VPIC) code (described in the next section) and the same computational time (9000 $\omega_{pe}^{-1}$). Our simulations only yielded weak $\rm H_2$ emission, without significant high-harmonic emissions whose generation mechanism thus remains to be revealed. This presents the main purpose of the present study.

\section{Numerical Setup and Parameters}

The simulations are performed using the VPIC code developed and released by Los Alamos National Labs, which is two-dimensional in space and three-dimensional for particle velocity and EM fields. VPIC employs a second-order, explicit, leapfrog algorithm to update charged particle positions and velocities in order to solve the relativistic kinetic equation for each species, along with a full Maxwell description for electric and magnetic fields evolved via a second-order finite-difference time-domain solver (\citealp{Bowers2008a,Bowers2008b, Bowers2009}). Periodic boundary conditions are used. The background magnetic field is set to be $ \vec{B}_0 (= B_0 \hat{e}_z) $, and the wavevector $ \vec{k} $ is in the $ xOz $ plane. The plasmas consist of three components, including background electrons and protons with a Maxwellian distribution and an electron beam with the following velocity distribution function:

\begin{equation} f_e=A_e\exp\left(-\frac{u_\perp^{2}}{2u_0^2}-\frac{(u_\parallel-u_d)^{2}}{2u_0^{2}}\right)
\label{E4.1}
\end{equation}
where $u_\parallel$ and $u_\perp$ are the parallel and perpendicular components of the momentum per mass, $u_d$ is the average drift momentum per mass of the beam electrons, $u_0$ is the thermal velocity of energetic electrons, and $A_e$ is the normalization factor.

Based on the observations of solar radio bursts (e.g., \citealp{Wild1950, Wild1959, Alvarez1973, Reid2014}) and in situ data of energetic electrons (e.g., \citealp{Lin1973, Lin1981, Lin1986}), the drift speed of the electron beam is set to be $u_d$ = 0.3--0.7$c$ and the thermal velocity of background and beam electrons with a fixed value of $u_0 = 0.028c$. The ratio of plasma oscillation frequency to electron gyrofrequency ($ \omega_{pe} / \Omega_{\rm ce} $) is set to be 100. The density ratio of beam-background electrons ($n_b/n_0$) is set to be $0.5\%$. All particles initially distribute homogeneously in space.

To facilitate the comparison with \cite{Krafft2022}, we use the same numerical configurations, such as the simulation domain ($L_x = L_z = 1024 \ \Delta$, where $\Delta = \sqrt{2} \ \lambda_{De}$ is the grid spacing and $\lambda_{De}$ is the Debye length of the background electrons), the unit of time ($\omega_{pe}^{-1}$), the simulation time 3000 $\omega_{pe}^{-1}$ and the time step $\Delta t = 0.01$ $\omega_{pe}^{-1}$. The corresponding resolvable range of $|k|$ is [0.15, 79] $\omega_{pe} / c $, and the range of $\omega$ is [0.002, 6.4] $\omega_{pe}$ (for the time interval of 3000 $\omega_{pe}^{-1}$). The NPPCPS is 1800 for background electrons and 900 for both background protons and beam electrons. The ratio of the ion to the electron plasma temperatures is set to be $T_i/T_e = 0.1$ to avoid strong Landau damping. At the start of the simulation, we maintain charge neutrality by setting proper weights to each species of macroparticle and neutralize the electron beam current by setting a proper bulk speed to the background electrons (see, e.g., \citealp{Henri2019, Chen2022, Zhang2022}).

\section{Numerical Results}
We first present the reference case (Case R) with beam velocity $u_d = 0.5c$ and the realistic proton-electron mass ratio ($m_i/m_e = 1836$) that will be compared with the corresponding thermal case (Case T) to tell the excitations of waves. Then, we vary $u_d$ to investigate its effect. One experiment with a larger mass ratio ($m_i/m_e = 18360$) is conducted to expand the investigation.

\subsection{Wave Analysis for Case R}
\label{sec:3.1}

\begin{figure*}
   \centerline{\includegraphics[width=0.9\textwidth]{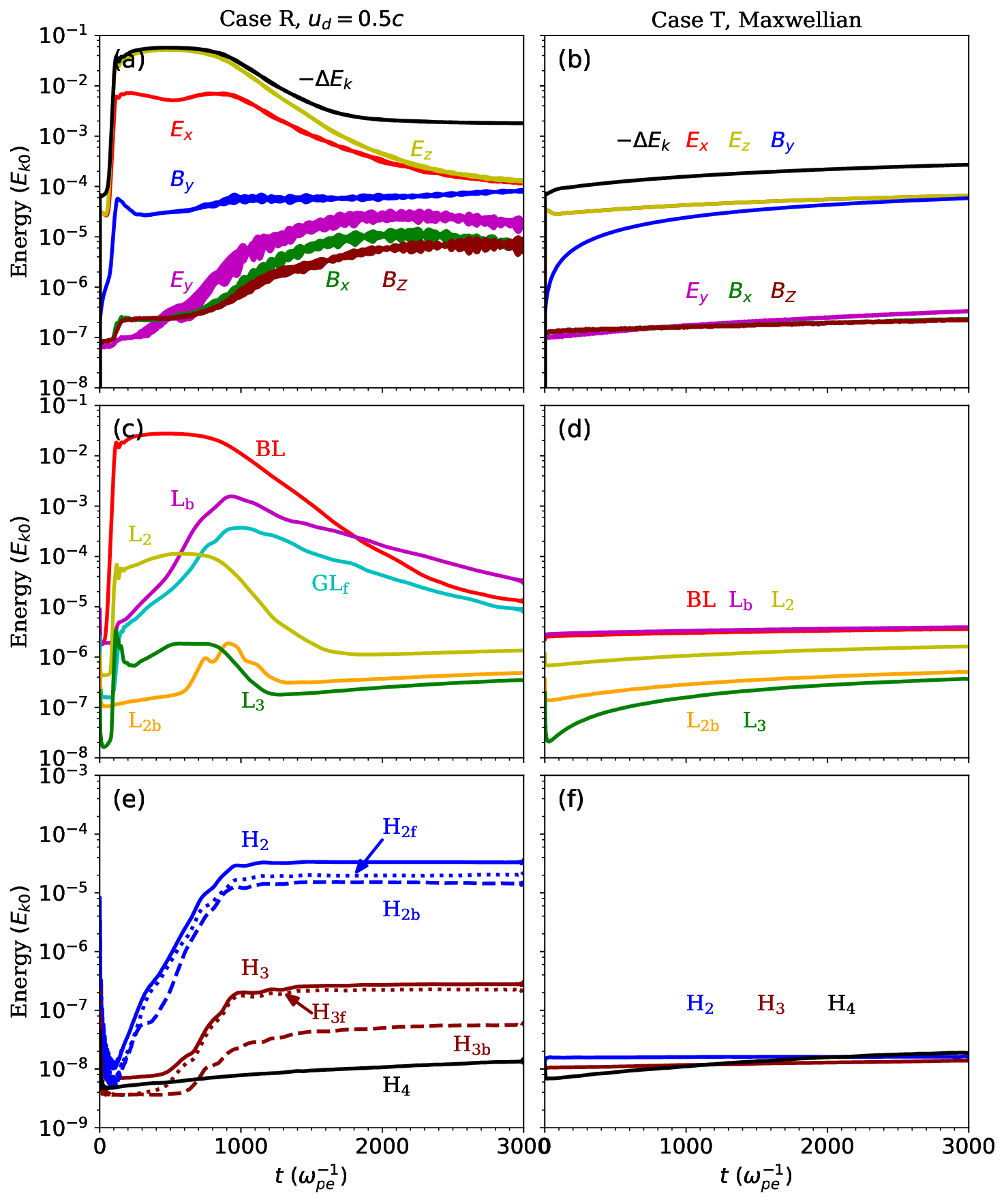}
              }
              \caption{
              (a) and (b) Temporal profiles of energies of various field components ($E_x, \ E_y, \ E_z, \ B_x, \ B_y$, and $B_z$) and the negative change of the total electron energy ($- \Delta E_k$). Temporal energy profiles of different modes, where (c) and (d) are for the ES modes and (e) and (f) are for the EM radiation. The subscripts ``f'' and ``b'' indicate the forward- and backward-propagating parts of the respective modes, while the subscript numbers represent the corresponding harmonics. The left column presents the results of Case R, and the right column displays the corresponding thermal Case T. The values in the plots are normalized by the respective initial kinetic energy of total electrons ($E_{k0}$).
              }
   \label{Fig:figure1}
   \end{figure*}

We first present the energy curves of the six field components and the negative change of the total kinetic energy of all electrons ($- \Delta E_k$) for Cases R and T (Figure \ref{Fig:figure1}(a) and (b)). The energies of all field components in Case R are significantly stronger than those in Case T by at least 2--4 orders of magnitude. This means these fields arise due to the beam-plasma interaction. Based on the energy profiles, the initial stage (0--100 $\omega_{pe}^{-1}$) is characterized by the rapid rise of $E_x$ and $E_z$ in energy, corresponding to the growth of the primary BL mode. At the end of this stage, $- \Delta E_k$ reaches $\sim$4$\%$ of $E_{k0}$. The maximum of $- \Delta E_k$ is $\sim$6$\%$ of $E_{k0}$, obtained at $\sim$300 $\omega_{pe}^{-1}$. From 100 to 900 $\omega_{pe}^{-1}$, the BL mode saturates at an energy level of 0.03 $E_{k0}$. This is followed by the damping stage, within which the wave energy is returned to electrons. Between 0 and 2000 $\omega_{pe}^{-1}$, the intensities of the other three components ($E_y$, $B_x$, and $B_z$) slowly rise and saturate. The $E_y$ component is mainly associated with the Z mode and the harmonic radiation, and the $B_x$ and $B_z$ components mainly carry the W mode, as well as the harmonic radiation.

\begin{figure*}
   \centerline{\includegraphics[width=0.9\textwidth]{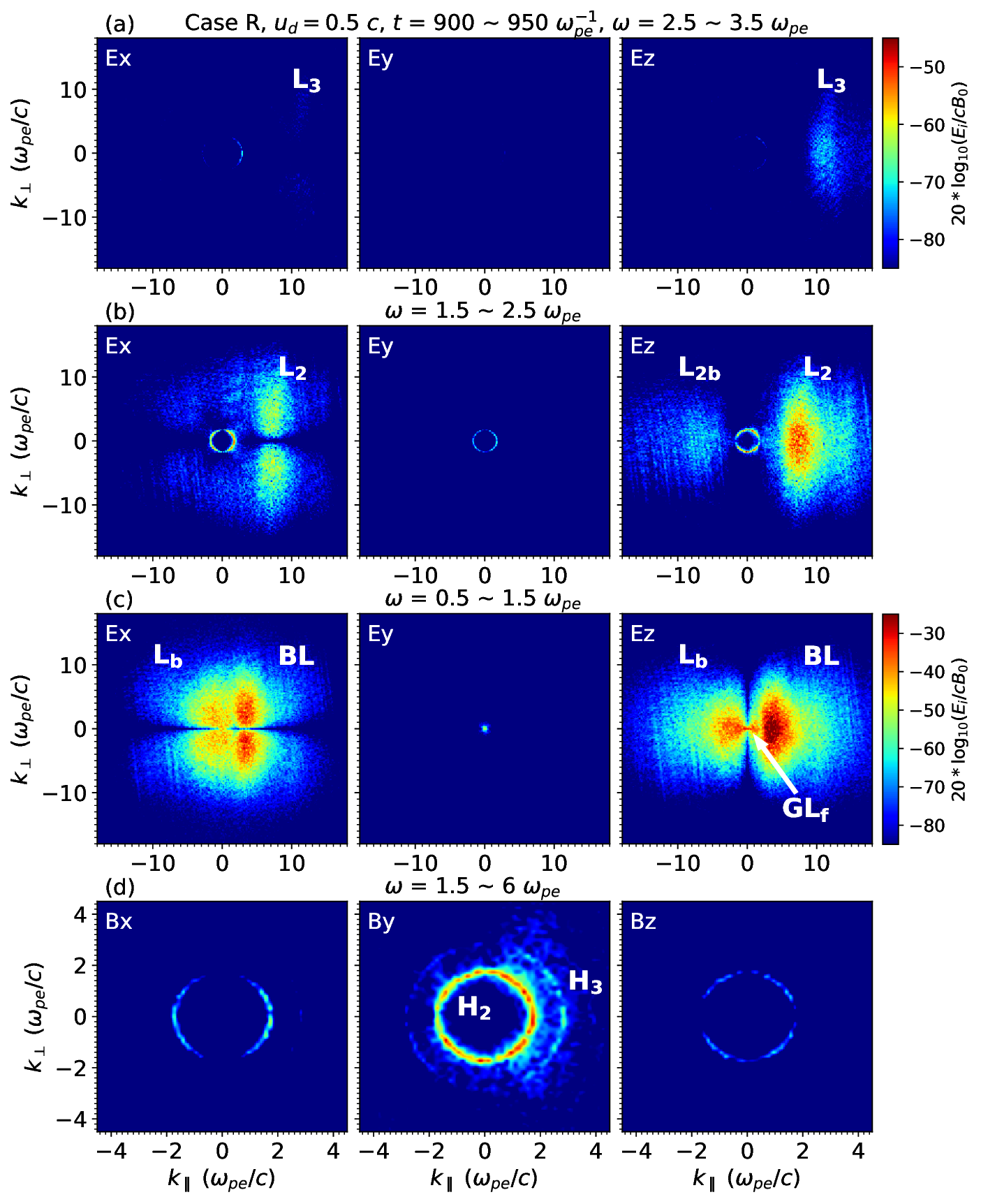}
              }
              \caption{
              Maximum intensity of the six field components in the $\omega$ domain ((a) 2.5--3.5, (b) 1.5--2.5, (c) 0.5--1.5, and (d) 1.5--6 $\omega_{pe}$) as a function of $k_\parallel$ and $k_\perp$ over the interval of $900 < \omega_{pe}t < 950$ for Case R. The upper three rows illustrate the corresponding ES modes, while the fourth row represents the EM radiation displayed by the magnetic field components. An animation of this figure is available. The animation begins with the map for the interval of [0, 50] $\omega_{pe}^{-1}$ and ends with that for the interval of [2950, 3000] $\omega_{pe}^{-1}$. The real-time duration of the video is 12 s.
              }
              {(An animation of this figure is available.)}
   \label{Fig:figure2}
   \end{figure*}

\begin{figure*}
   \centerline{\includegraphics[width=0.9\textwidth]{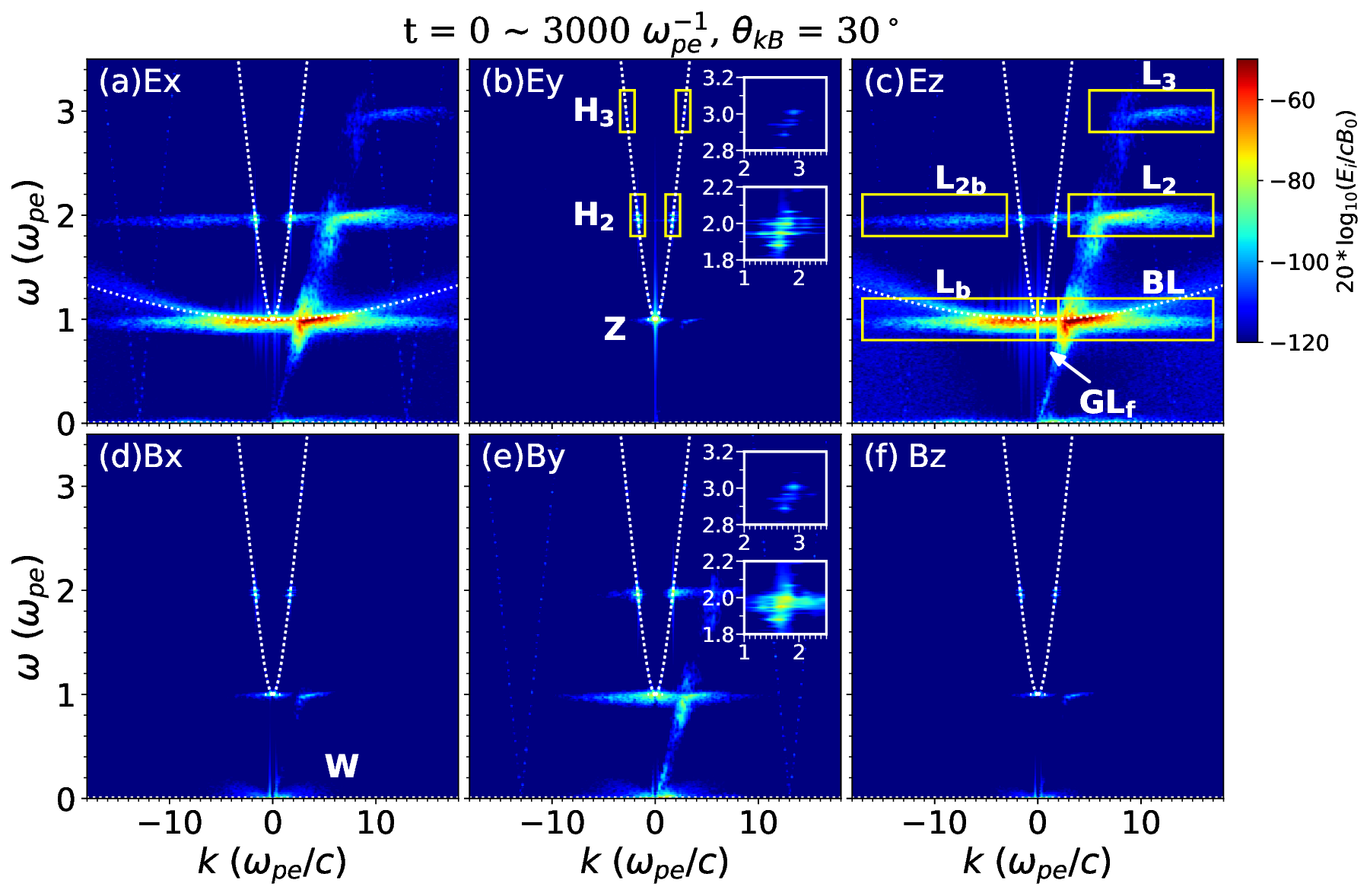}
              }
              \caption{
              Wave dispersion diagrams of the six field components along $\theta_{kB} = 0^\circ$ ($\theta_{kB}$ is the angle between the wavevector, $ \vec{k} $, and the background magnetic field, $ \vec{B}_0 $). The curve near $\omega_{pe}$ in panel (c) represents the dispersion relation of L wave with thermal effects ($\omega^2 = \omega_{pe}^2 (1+3k^2 \lambda_{De}^2)$), while the other white dashed lines represent the dispersion curves given by the magnetoionic theory. ``Z'' denotes the Z mode, and ``W'' denotes the whistler mode. The yellow boxes mark the regions selected for the evaluation of mode energy. The white boxes in panels (b) and (e) zoom in on the $\rm H_2$ and $\rm H_3$ spectra. An animation of this figure is available. The animation begins at $\theta_{kB} = 0^\circ$ and advances $10^\circ$ at a time up to $\theta_{kB} = 90^\circ$. The real-time duration of the video is 3 s.
              }
              {(An animation of this figure is available.)}
   \label{Fig:figure3}
   \end{figure*}

Figure \ref{Fig:figure2} presents wave-energy maps in the wavevector ($\vec{k}$) space for the ES modes around $\omega_{pe}$, 2 $\omega_{pe}$, and 3 $\omega_{pe}$ and the harmonic radiations. The maps present the intensity maxima of waves at the corresponding $\vec{k}$, regardless of their frequencies. The mode nature can be identified with the analytic dispersion curves plotted in Figure \ref{Fig:figure3} (and the accompanying movies). The modes excited here include the $\rm L_3$ mode, the $\rm L_2$ and $\rm L_{2b}$ modes, the BL and $\rm L_b$ modes, and the $\rm H_2$ and $\rm H_3$ emissions. Here, the subscript ``f'' (``b'') denotes the forward (backward)-propagating portion of the corresponding mode, while the number in the subscript indicates the harmonics. The dominant $E_x$ and $E_z$ components correspond to the primary forward-propagating ES-BL mode (Figure \ref{Fig:figure2}(c)). The BL mode extends $\sim$$\pm 77^\circ$  away from the parallel direction in $E_z$, with a significant perpendicular electric field component of $E_x$. The movie accompanying Figure \ref{Fig:figure2} shows its evolution in the $\vec{k}$ space. During the rapid-excitation stage, the $E_z$ component presents a rather narrow range in the $\vec{k}$ space, with $1.5 \leq k_\parallel (\omega_{pe}/c) \leq3.5$ and $k_\perp (\omega_{pe}/c) \leq 8$; later, it expands toward larger $k_\parallel$. It is within the range of $2 \leq k_\parallel (\omega_{pe}/c) \leq 15$ at $\sim$900 $\omega_{pe}^{-1}$, followed by the decline of intensity and the shrinkage of the $\vec{k}$ range.

Figures \ref{Fig:figure2}(c) and \ref{Fig:figure3}(c) reveal the presence of two additional L components on the left (backward) side of the BL mode, referred to as the forward- and backward-propagating generalized Langmuir (GL) modes, represented by $\rm GL_f$ and $\rm L_b$, respectively (see also \citealp{Chen2022, Zhang2022}). As depicted in Figure \ref{Fig:figure2} and its accompanying movie, both the $\rm GL_f$ and $\rm L_b$ modes experience delayed growth with weaker intensities compared to the BL mode. The $k_\parallel$ range of the $\rm GL_f$ mode remains relatively narrow ($0 \leq k_\parallel (\omega_{pe}/c) \leq 2$), while the $k_\parallel$ of the $\rm L_b$ mode expands gradually and reaches $k_\parallel(\omega_{pe}/c) \sim $ -15 at $\sim$900 $\omega_{pe}^{-1}$, with an angular pattern resembling that of the primary BL mode. Subsequently, both modes damp with decreasing ranges of $\vec{k}$.

The intensities of the BL, $\rm L_2$, and $\rm L_3$ modes decrease gradually with increasing ranges of $k_\parallel$. The $k_\parallel$ range of $\rm L_2$ expands from [3, 6] $\omega_{pe}/c$ initially to [5, 15] $\omega_{pe}/c$ around 900 $\omega_{pe}^{-1}$. For $\rm L_3$, its $k_\parallel$ range extends from [6, 8] to [9, 13] $\omega_{pe}/c$ during [100, 900] $\omega_{pe}^{-1}$. The $\rm L_{2b}$ mode appears around 650 $\omega_{pe}^{-1}$, within $-10 \leq k_\parallel(\omega_{pe}/c) \leq -4$. After $\sim$900 $\omega_{pe}^{-1}$, these ES waves begin to decay with the shrinkage of the $k_\parallel$ range.

The circular patterns shown in Figure \ref{Fig:figure2}(d) represent the $\rm H_2$ and $\rm H_3$ radiation. We observe no signatures of higher-harmonic emissions. According to the movie accompanying Figure \ref{Fig:figure2}, the $\rm H_3$ emission grows after $\rm H_2$. The $\rm H_2$ radiation is stronger than $\rm H_3$, and the forward-propagating portion of $\rm H_3$ is stronger than the backward-propagating portion.

In Figure \ref{Fig:figure3}, we present the $\omega-k$ dispersion analysis along the propagation angle $\theta_{kB} = 30^\circ$. The accompanying movie shows the variation of the dispersion relation at a step of $10^\circ$. Note that the frequency ranges of $\rm H_2$ and $\rm H_3$ vary similarly with $\theta_{kB}$. The overplotted dashed lines represent the dispersion relations of the L wave with thermal effects ($\omega^2 = \omega_{pe}^2 (1+3k^2 \lambda_{De}^2)$) and wave modes given by the classical cold plasma magnetoionic theory. Table \ref{Tab:Table1} summarizes the $\omega$ and $k$ ranges for each mode just before the rapid decay of the ES modes ($\sim$900 $\omega_{pe}^{-1}$).

We examine the temporal development of various wave modes with energy profiles for specific field component(s) (see Figures \ref{Fig:figure1}(c) and (e)), which are calculated by integrating the field energy within a specified spectral range (Figure \ref{Fig:figure3}(c)) along the corresponding dispersion curve according to Parseval's theorem. The forward-propagating BL, $\rm L_2$, and $\rm L_3$ modes display the strongest and fastest growth; their intensities saturate around 100 $\omega_{pe}^{-1}$, then remain at an almost constant level before declining gradually after $\sim$900 $\omega_{pe}^{-1}$. The $\rm L_b$ mode is slightly stronger than $\rm GL_f$, both with similar energy profiles. Both modes grow gradually over time, reaching the maximum level near 900 $\omega_{pe}^{-1}$. The $\rm L_{2b}$ mode shows a delayed onset of growth at $\sim$600 $\omega_{pe}^{-1}$, followed by further enhancement and subsequent decay to the thermal noise level.

The comparison with the corresponding thermal case (Figures \ref{Fig:figure1}(e) and (f)) confirms the significant enhancement of the $\rm H_2$ and $\rm H_3$ emissions over thermal noise, while emissions at the fourth and higher harmonics are absent.

The energy profiles of $\rm H_2$ and $\rm H_3$ are similar. Both modes get enhanced gradually over time, reaching the saturation level near 900 $\omega_{pe}^{-1}$ and remaining at a nearly constant level thereafter. However, the onset of $\rm H_3$ is later by $\sim$ 400 $\omega_{pe}^{-1}$ compared to $\rm H_2$. The $\rm H_3$ intensity gets weaker than $\rm H_2$, and their intensity ratio ($\rm W_{H_2}/W_{H_3}$) at saturation is $\sim$120.

The forward- (f) and backward- (b) propagating portions of $\rm H_2$ show little difference, and the intensity ratio ($\rm W_{H_{2f}}/W_{H_{2b}}$) at saturation is $\sim$1.4. However, the forward-propagating portion of $\rm H_3$ is much stronger than its backward-propagating portion, with $\rm W_{H_{3f}}/W_{H_{3b}}$ $\sim$ 4 at saturation, agreeing with the wave-energy maps shown in Figure \ref{Fig:figure2}(d).

\subsection{Effect of the Beam Velocity and the Mass Ratio ($m_i/m_e$)}

Two sets of numerical experiments are carried out. We first investigate the effect of the beam velocity ($u_d$) with the physical mass ratio ($m_i/m_e$ = 1836). This gives four additional cases (A, B, C, and D), corresponding to $u_d = 0.3c, \ 0.4c, \ 0.6c$, and $0.7c$, respectively. We then investigate the effect of $m_i/m_e$, since this parameter can affect the intensity of the ES modes and the radiation process (\citealp{Chen2022, Zhang2022}). This gives Case E with $u_d = 0.5c$ and $m_i/m_e$ = 18360.

\begin{figure*}
   \centerline{\includegraphics[width=0.9\textwidth]{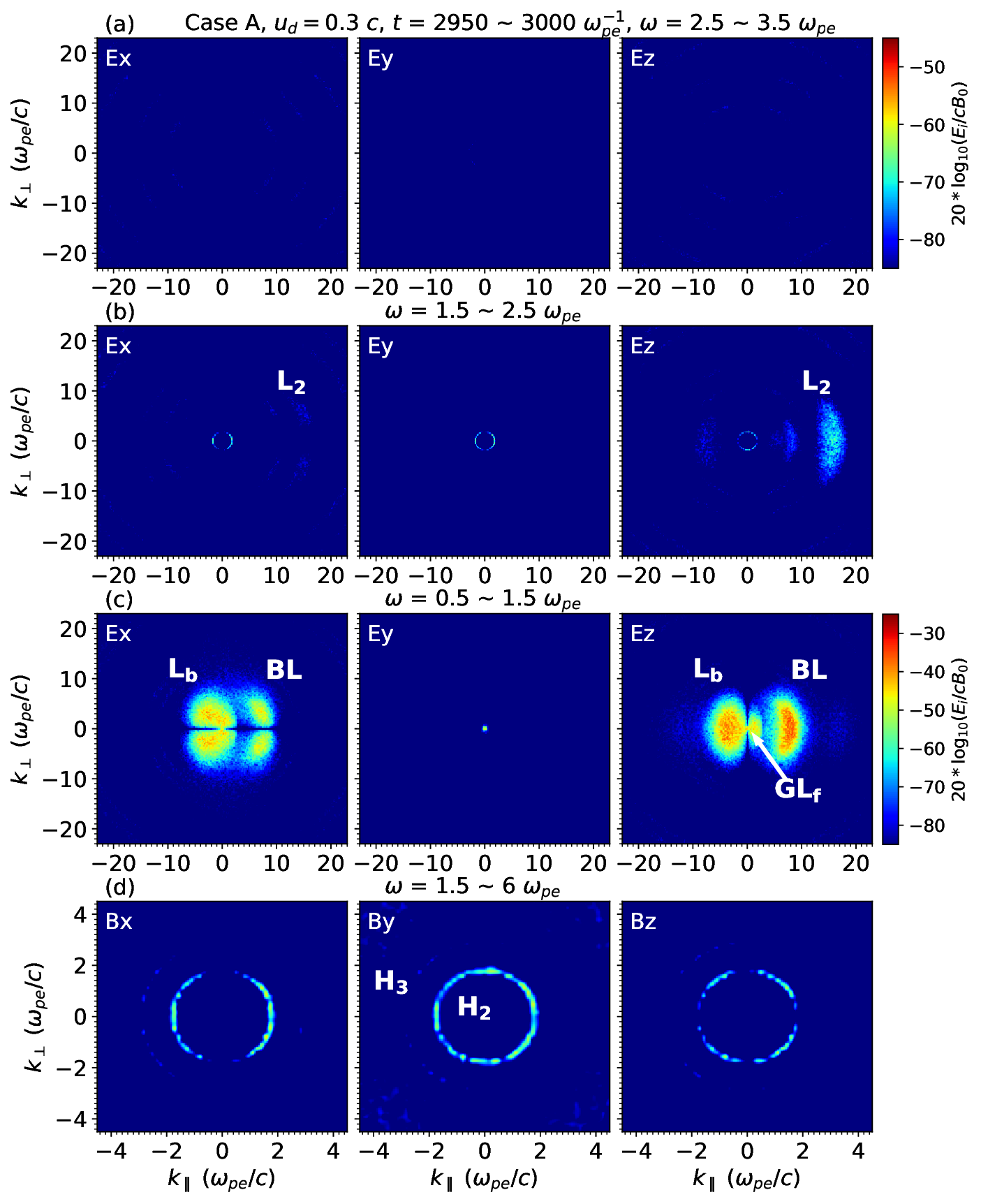}
              }
              \caption{
              Maximum intensity of the six field components in the $\omega$ domain ((a) 2.5--3.5, (b) 1.5--2.5, (c) 0.5--1.5, and (d) 1.5--6 $\omega_{pe}$) as a function of $k_\parallel$ and $k_\perp$ over the interval of $2950 < \omega_{pe}t < 3000$ for Case A. The upper three rows illustrate the corresponding ES modes, while the fourth row represents the EM radiation displayed by the magnetic field components. An animation of this figure is available, and it presents the mode evolution for Cases A, B and E, with each part beginning with the map for the interval of [0, 50] $\omega_{pe}^{-1}$ and ending with that for the interval of [2950, 3000] $\omega_{pe}^{-1}$. The real-time duration of the video is 36 s.
              }
              {(An animation of this figure is available.)}
   \label{Fig:figure4}
   \end{figure*}

\begin{figure*}
   \centerline{\includegraphics[width=0.9\textwidth]{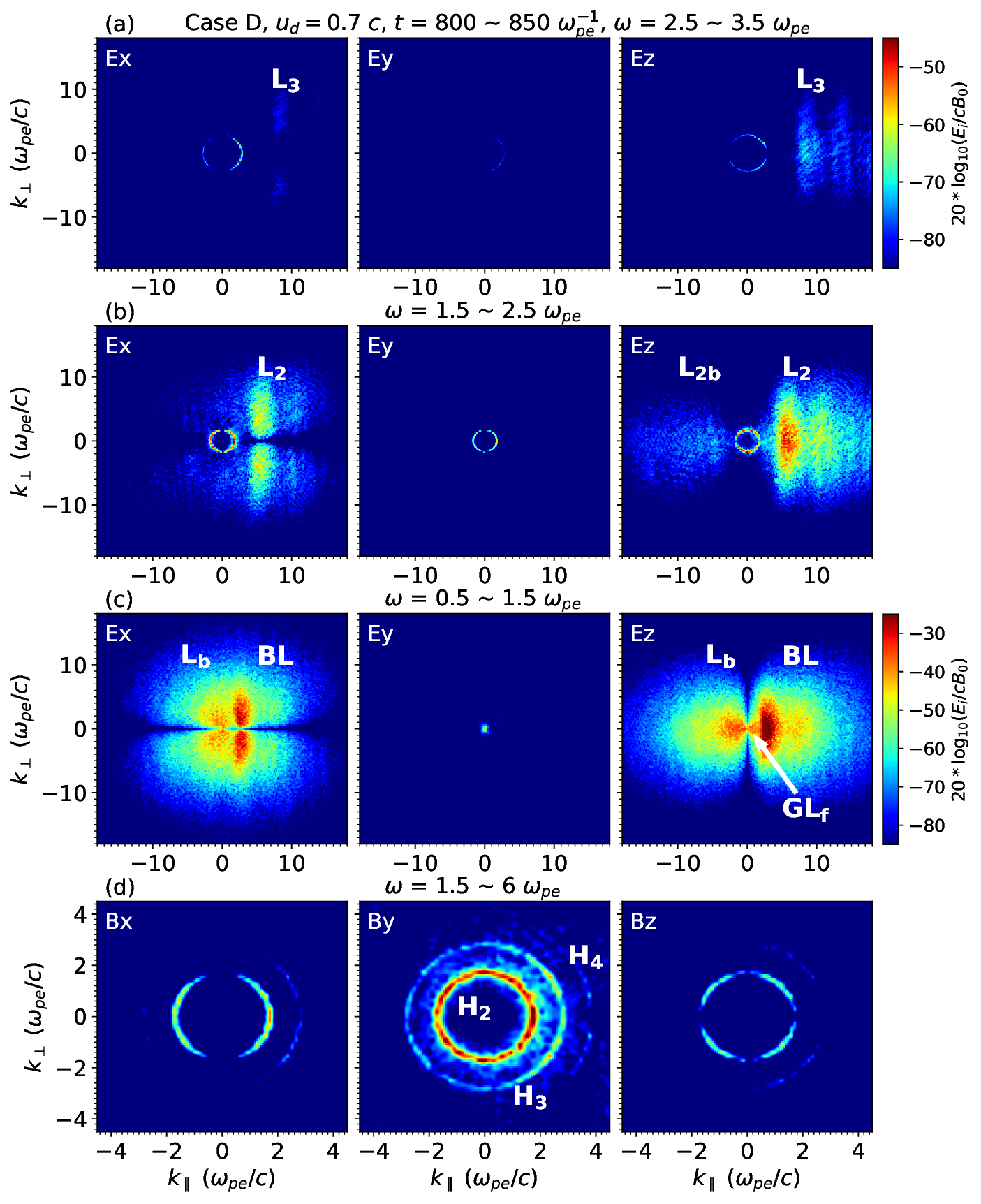}
              }
              \caption{
              Maximum intensity of the six field components in the $\omega$ domain ((a) 2.5--3.5, (b) 1.5--2.5, (c) 0.5--1.5, and (d) 1.5--6 $\omega_{pe}$) as a function of $k_\parallel$ and $k_\perp$ over the interval of $800 < \omega_{pe}t < 850$ for Case D. The upper three rows illustrate the corresponding ES modes, while the fourth row represents the EM radiation displayed by the magnetic field components. The video begins with the map for the interval of [0, 50] $\omega_{pe}^{-1}$, ending with that for the interval of [2950, 3000] $\omega_{pe}^{-1}$. The real-time duration of the video is 12 s.
              }
              {(An animation of this figure is available.)}
   \label{Fig:figure5}
   \end{figure*}

Figures \ref{Fig:figure4} and \ref{Fig:figure5} present the wave-energy maps in the $\vec{k}$ space for Cases A ($u_d=0.3c$) and D ($u_d=0.7c$), respectively. The accompanying movies show the evolution. In comparison with Case R ($u_d=0.5c$), the forward-propagating ES modes ($\rm BL$, $\rm L_2$ and $\rm L_3$) have larger $k_\parallel$ values in Case A, but smaller values in Case D. For instance, the $k_\parallel$ range of $\rm BL$ in Case A extends from [3, 6.5] $\omega_{pe}/c$ initially to [4, 11] $\omega_{pe}/c$, while it expands from [1, 2] $\omega_{pe}/c$ to [1.5, 15] $\omega_{pe}/c$ in Case D.

For EM harmonics, the $\rm H_2$ and $\rm H_3$ emissions exhibit weaker intensities in Case A than in Case R, with $\rm H_3$ only presenting a faint backward-propagating part ($\rm H_{3b}$; see Figure \ref{Fig:figure4}(d) and the accompanying movie). Figure \ref{Fig:figure5}(d) shows the significant enhancements of $\rm H_2$ and $\rm H_3$, with the additional appearance of $\rm H_4$ emission in the forward-propagation direction ($\rm H_{4f}$). According to the movie accompanying Figure \ref{Fig:figure5}, the $\rm H_3$ emission grows after $\rm H_2$, and $\rm H_4$ grows after $\rm H_3$.

\begin{figure*}
   \centerline{\includegraphics[width=0.9\textwidth]{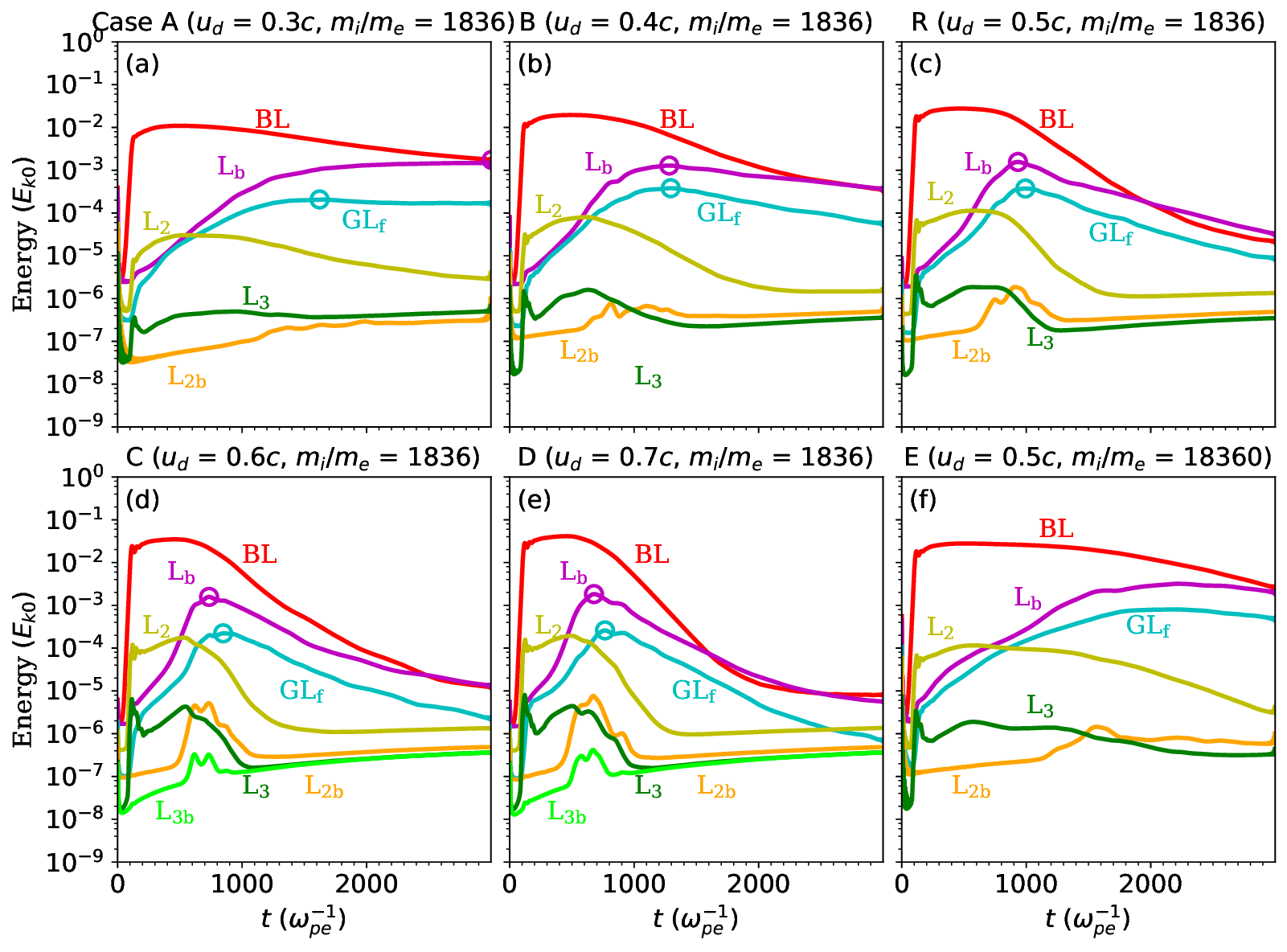}
              }
              \caption{
               The temporal energy profiles of the ES harmonic modes (BL, $\rm L_b, \ GL_f, \ L_2, \ L_{2b}, \ L_3, \ L_{3b}$), normalized by the respective initial kinetic energy of total electrons ($E_{k0}$). (a)--(e) are for cases with $m_i/m_e$ = 1836 and $u_d$ = 0.3--0.7$c$, and (f) is for Case E with $u_d = 0.5c$ and $m_i/m_e = 18360$. The magenta and cyan circles in panels (a)--(e) mark the intensity maxima of the $\rm L_b$ and $\rm GL_{f}$ modes, respectively.
              }
   \label{Fig:figure6}
   \end{figure*}

Figure \ref{Fig:figure6} illustrates the temporal energy profiles of the ES modes for different cases. As $u_d$ increases from $0.3c$ to $0.7c$, the intensity maximum of each ES mode increases correspondingly. Among them, the BL, $\rm L_2$, and $\rm L_3$ modes exhibit a similar trend, with the onset of the decay being earlier for larger $u_d$. The $\rm L_b$ and $\rm GL_f$ modes arise simultaneously and both grow and decay faster with increasing $u_d$. Both modes reach their intensity maximum earlier with increasing $u_d$. The times of their intensity maxima are indicated by circles in Figure \ref{Fig:figure6} at 3000, 1282, 933, 735, and 677 $\omega_{pe}^{-1}$ for $\rm L_b$ mode and 1622, 1294, 993, 848, and 766 $\omega_{pe}^{-1}$ for $\rm GL_f$ mode. These times correspond to the start of the decay of the respective BL mode.

Figure \ref{Fig:figure7} illustrates the energy profiles of $\rm H_2$ and $\rm H_3$. Both show considerable enhancements in Cases B--E and R, while significant $\rm H_4$ emission appears only in Cases C and D, corresponding to $u_d = 0.6c$ and $0.7c$. For different cases, the energy profiles of $\rm H_2$, $\rm H_3$, and $\rm H_4$ are quite similar, with comparable growth rates and saturation times. The $\rm H_2/H_3$ emissions saturate earlier with increasing $u_d$. The $\rm H_2$ exhibits higher intensity and grows earlier than $\rm H_3$. The $\rm H_2/H_3$ energy ratio ($\rm W_{H_2}/W_{H_3}$) decreases with increasing $u_d$. For instance, as $u_d$ increases from $0.3c$ to $0.7c$, $\rm W_{H_2}/W_{H_3}$ decreases from 206.8 to 44.1; see Table \ref{Tab:Table2} for details. Likewise, $\rm H_3$ exhibits higher intensity and grows earlier than $\rm H_4$. In Cases C and D, the $\rm H_3/H_4$ energy ratios ($\rm W_{H_3}/W_{H_4}$) are 93.6 and 88.5, respectively.

The $\rm H_{2f}$ portion is stronger than $\rm H_{2b}$ for smaller $u_d$ (= $0.3c$ and $0.4c$), while $\rm H_{2f}$ is closer to $\rm H_{2b}$ in intensity with larger $u_d$ (from $0.5c$ to $0.7c$). Regarding $\rm H_3$, the $\rm H_{3b}$ portion is stronger than $\rm H_{3f}$ with $u_d = 0.3c$ (Case A). As $u_d$ increases to $0.4c$ (Case B), $\rm H_{3b}$ is initially weaker, yet it gets stronger than $\rm H_{3f}$ later. For $u_d \geq 0.5c$, $\rm H_{3f}$ is always stronger than $\rm H_{3b}$. In Cases C and D, $\rm H_{4f}$ is always stronger than $\rm H_{4b}$.

\subsection{On the Generation Mechanism of $\rm H_3$}
As introduced, three nonlinear wave-wave coupling processes have been proposed to account for the $\rm H_3$ emission. The modes involved must satisfy the matching conditions: $\vec{k}_1  + \vec{k}_2 = \vec{k}_3$ and $\omega_1 + \omega_2 = \omega_3$. For Case R with $u_d = 0.5c$, according to the $\omega$ and $k$ ranges of various modes listed in Table \ref{Tab:Table1}, three processes can satisfy the matching conditions: (a) $\rm BL + L_{2b} \rightarrow H_3$, (b) $\rm L_b + L_2 \rightarrow H_3$, and (c) $\rm BL/L_b + H_2 \rightarrow H_3$.

These possibilities can be further evaluated by analyzing the energy profiles of relevant modes. From Figures \ref{Fig:figure1}(c) and (e), we have (1) the energy curves of $\rm L_b$, $\rm H_2$, and $\rm H_3$ are correlated with similar growth rates and consistent saturation times ($\sim$900 $\omega_{pe}^{-1}$); (2) $\rm H_2$ exhibits higher intensity and grows earlier than $\rm H_3$; (3) $\rm L_{2b}$ is much weaker than other modes and damps to the noise level shortly after the brief enhancement; and (4) $\rm H_3$ grows earlier than $\rm L_{2b}$.

The first point suggests that $\rm L_b$ may play a role in the generation of both $\rm H_2$ and $\rm H_3$. Combining the first two points, $\rm H_2$ may get involved in generating $\rm H_3$, which supports process (c).

The third and forth points rule out the participation of $\rm L_{2b}$ in generating $\rm H_3$, i.e., process (a).

Process (b) ($\rm L_b + L_2 \rightarrow H_3$) can be largely rejected by considering the late-stage (say, after $t$ = 2000 $\omega_{pe}^{-1}$) characteristics of $\rm H_3$ and $\rm L_2$ in Cases A, B, and E. We observe from Figures \ref{Fig:figure7}(a), (b), and (f) that $\rm H_3$ presents persistent enhancement during this late stage. Yet, according to Figures \ref{Fig:figure6}(a), (b), and (f) and the movies accompanying Figure \ref{Fig:figure4}, in the late stage, (1) in Cases A and B, $\rm L_2$ damps significantly and approaches its initial level of numerical noise; and (2) in Case A the $k_\parallel$ of $\rm L_2$ shifts from the earlier range of $\sim$11--17 $\omega_{pe}/c$ to $\sim$13--17 $\omega_{pe}/c$, while that of $\rm L_b$ remains in the range of -6--0 $\omega_{pe}/c$, making it impossible to meet the corresponding matching condition of process (b). A similar shift occurs to Case E in which the $k_\parallel$ of $\rm L_2$ shifts from the earlier range of $\sim$6--11 to $\sim$9--12 $\omega_{pe}/c$ while that of $\rm L_b$ remains in the range of -7--0 $\omega_{pe}/c$, making it difficult to meet the corresponding matching condition for the generation of $\rm H_{3b}$ with process (b). We conclude that these observations do not favor the significance of process (b) in generating $\rm H_3$.

Thus, the most likely process to generate $\rm H_3$ is (c), $\rm BL/L_b + H_2 \rightarrow H_3$. Based on the matching conditions of three-wave interaction, we can further separate the process into two subprocesses: $\rm BL + H_{2b} \rightarrow H_{3f}$ and $\rm L_b+ H_{2f} \rightarrow H_{3b}$. Note that $\rm H_{3f}$ is stronger and rises earlier than $\rm H_{3b}$, indicating that the process $\rm BL + H_{2b} \rightarrow H_{3f}$ is more effective.

We observe a different directional pattern of $\rm H_3$ in different cases (A, B, and R). This can be used to constrain the generation mechanism of $\rm H_{3f}$ and $\rm H_{3b}$. According to Figure \ref{Fig:figure4} and the accompanying movie, the $k$ value of the BL mode ($k_{\rm BL}$) is too large in Case A ($u_d = 0.3c$) to coalesce with $\rm H_2$ so as to generate $\rm H_{3f}$; this explains the absence of $\rm H_{3f}$ in this case. On the other hand, $\rm L_b$ has the appropriate wavenumber to coalesce with $\rm H_2$ and generate $\rm H_{3b}$. For Case B with $u_d = 0.4c$ (see the second part of the accompanying movie), the value of $k_{\rm BL}$ allows the occurrence of the process $\rm BL + H_{2b} \rightarrow H_{3f}$. Note that $k_{\rm BL}$ increases with increasing time; this makes it more difficult to meet the corresponding matching conditions, while the conditions for $\rm L_b + H_{2f} \rightarrow H_{3b}$ can still be satisfied. This can explain the relative intensity variation of $\rm H_{3b}$ and $\rm H_{3f}$. For Case R with $u_d = 0.5c$ (see Figure \ref{Fig:figure2} and the accompanying movie), both the BL and $\rm L_b$ modes can coalesce with $\rm H_2$ during the simulation. We get a stronger $\rm H_{3f}$ than $\rm H_{3b}$ since BL is always stronger than $\rm L_b$.

Note that in Case A we observed the simultaneous presence of $\rm GL_f$ and $\rm H_2$, yet the $\rm H_{3f}$ radiation is always at the noise level. This indicates that the process of $\rm GL_f + H_2 \rightarrow H_3$ is not important here.

\subsection{On the Generation Mechanism of $\rm H_4$}
In the following, we discuss the generation mechanism of $\rm H_4$. Based on the matching conditions, the following processes are possible: (a) $\rm BL + L_{3b} \rightarrow H_4$, (b) $\rm L_b + L_3 \rightarrow H_4$, (c) $\rm BL/L_b + H_3 \rightarrow H_4$, and (d) $\rm L_2/L_{2b} + H_2 \rightarrow H_4$.

\begin{figure*}
   \centerline{\includegraphics[width=0.9\textwidth]{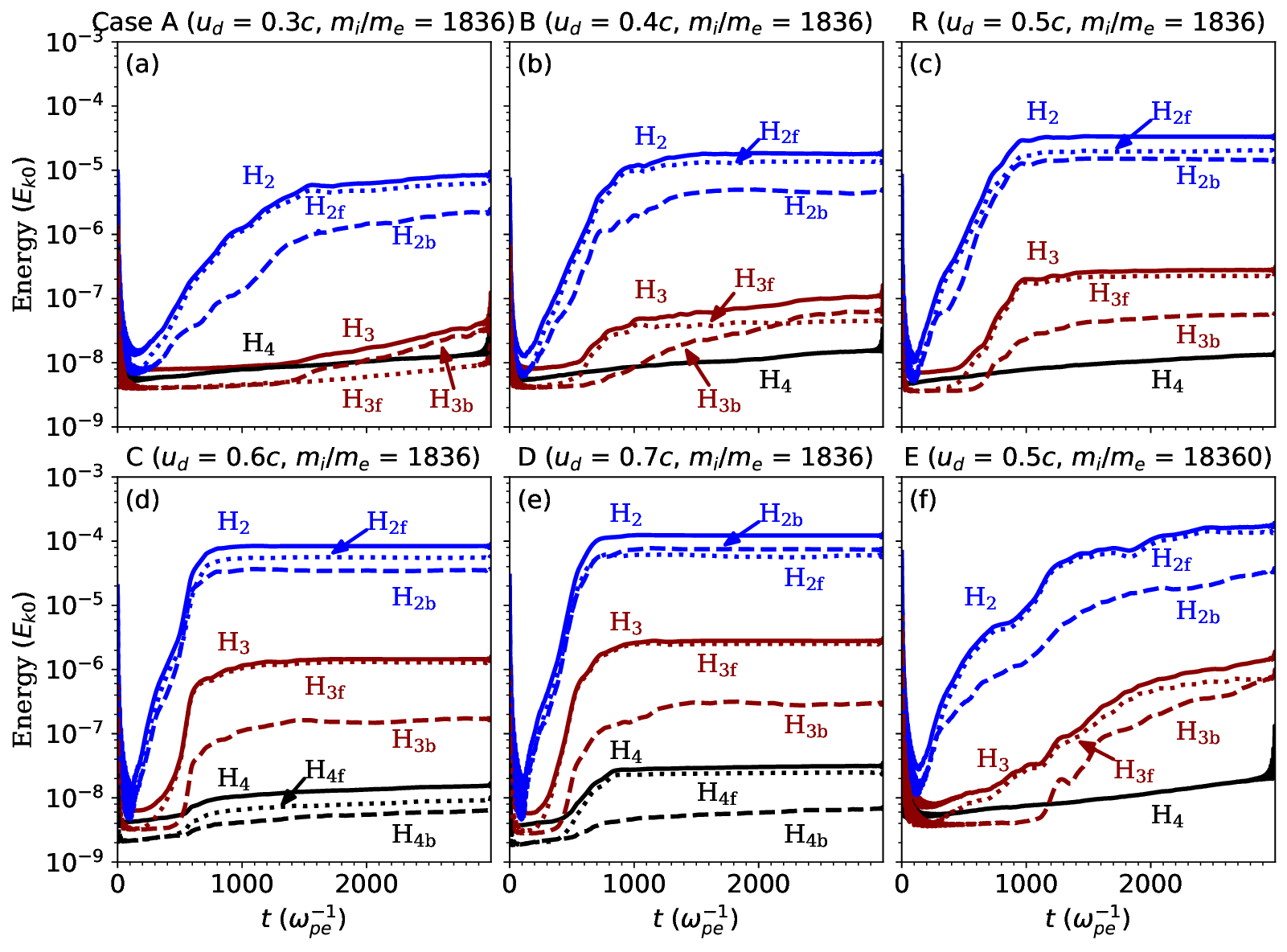}
              }
              \caption{
               The temporal energy profiles of the second, third, and fourth EM harmonic emission. The solid lines represent the sum of the energy in each direction of the corresponding harmonic emission, and the dotted and dashed lines represent the forward (f; $\ k_\parallel  > 0$) and backward (b; $\ k_\parallel < 0$) propagation part of the energy of the corresponding radiation, respectively. The values in the figures are normalized by the respective initial kinetic energy of total electrons ($E_{k0}$).
              }
   \label{Fig:figure7}
   \end{figure*}

First, we can exclude process (a) involving the $\rm L_{3b}$ mode. According to Figures \ref{Fig:figure6}(d) and (e) and \ref{Fig:figure7}(d) and (e), during the initial growth of $\rm H_4$ $\rm L_{3b}$ is at the noise level before 600 $\omega_{pe}^{-1}$ and then presents a local peak from 600 to 800 $\omega_{pe}^{-1}$. Therefore, it is unlikely to be important for the radiation of $\rm H_4$.

We can also exclude process (b) based on the following two aspects. First, according to Figures \ref{Fig:figure6}(d) and (e) and \ref{Fig:figure7}(d) and (e), $\rm L_b$ first increases and then decreases in intensity along with the $\rm H_4$ growth, while $\rm L_3$ exhibits a decreasing trend in intensity. This is different from the $\rm H_4$ intensity profile. Second, $\rm L_3$ does not change much in intensity when $u_d$ increases from $0.6c$ to $0.7c$, while $\rm H_4$ reveals significant enhancement.

We can further exclude process (d). From Figure \ref{Fig:figure7}(e), $\rm H_3$ starts to grow earlier than $\rm H_4$. As stated earlier, $\rm H_2$ contributes to the $\rm H_3$ radiation. If it also contributes to the $\rm H_4$ radiation, e.g., via process (d), then $\rm H_3$ and $\rm H_4$ shall rise at a close time. This is inconsistent with our simulation. In addition, when $u_d$ increases from $0.6c$ to $0.7c$, $\rm L_{2b}$ does not change much in intensity, while $\rm H_4$ presents substantial enhancement (see Figures \ref{Fig:figure7}(d) and (e)). This indicates that the process involving $\rm L_{2b}$ in (d) can be ruled out.

Thus, we are left with process (c), with $\rm BL/L_b + H_3 \rightarrow H_4$. This is supported by the similar intensity profiles of $\rm L_b$, $\rm H_3$, and $\rm H_4$. If further taking the matching conditions into account, we find that the following subprocesses may act in the system: $\rm BL + H_{3b} \rightarrow H_{4f}$ and $\rm L_b+ H_{3f} \rightarrow H_{4b}$.

\section{Conclusions and Discussion}
In this study, we conduct fully EM PIC simulations to investigate the interaction between beam electrons and a weakly magnetized plasma with the solar coronal-interplanetary conditions. The purpose is to reveal the generation mechanism of high-harmonic plasma radiations. We obtained significant second, third, and fourth harmonic ($\rm H_2$, $\rm H_3$, and $\rm H_4$) radiations. The neighboring harmonics are different in intensity by about 2 orders of magnitude, with higher harmonics being weaker. Cases with larger beam velocities tend to favor the generation of higher harmonics and stronger radiation for the specific harmonic, which is consistent with the statement of \cite{Rhee2009}, ``by increasing the beam velocity, the present case allows for the excitation of higher-harmonic modes.'' We suggest that the $\rm H_3$ radiation is mainly generated via the process of $\rm H_2 + BL/L_b \rightarrow H_3$ and the $\rm H_4$ radiation is mainly generated via the process of $\rm H_3 + BL/L_b \rightarrow H_4$, with the forward portion given by the coalescence of $\rm H_2$ ($\rm H_3$) with BL mode and the backward portion given by its coalescence with $\rm L_b$.

The obtained results are helpful to understand the data of solar radio bursts with high-harmonic emission. For instance, the presence of $\rm H_3$ emission in observations suggests that the energetic electron beam has a higher-than-usual velocity (say, $>\sim0.5c$), and the intensity ratio of $\rm H_3$ and $\rm H_2$ depends on the beam velocity, with a larger intensity ratio corresponding to a faster beam.

According to our earlier simulations (\citealp{Chen2022, Zhang2022}) that investigated the PE process for weakly magnetized plasmas with $\omega_{pe} / \Omega_{\rm ce} = 10$ and nonmagnetized plasmas (i.e., $\omega_{pe}/\Omega_{ce} \sim \infty$), the excitation of primary modes and escaping emissions is quite similar for these two cases. Thus, we expect that the major conclusions deduced here still hold for other weakly magnetized plasmas as long as the values of $\omega_{pe} / \Omega_{\rm ce}$ are much larger than unity.

\section*{Acknowledgements}
This study is supported by NNSFC grants (Nos. 12103029, 1973031, 12203031), a NSFSP (Natural Science Foundation of Shandong Province) grant (No. ZR2021QA033) and the China Postdoctoral Science Foundation (No. 2021M691904). The authors acknowledge the National Supercomputer Centers in Tianjin and the Beijing Super Cloud Computing Center (BSCC; http://www.blsc.cn/) for providing high-performance computing (HPC) resources, and the open-source Vector-PIC (VPIC) code provided by Los Alamos National Labs (LANL). The authors are grateful to the anonymous referee for valuable comments.


\begin{table}[htbp]
\centering
\caption{Ranges of $\omega$ and $k$ for each mode at the moment before the rapid decay of each ES harmonic mode ($\sim$900 $\omega_{pe}^{-1}$) for Case R.}
\setlength{\tabcolsep}{5mm}{
\begin{tabular}{ccccccccc}
\hline
\hline
                       & BL       & $\rm L_b$  & $\rm GL_f$ & $\rm L_2$ & $\rm L_{2b}$ & $\rm L_3$ & $\rm H_2$ & $\rm H_3$  \\
\hline
$\omega \ (\omega_{pe})$ & 0.8--1.18 & 0.88--1.04  & 0.9--1      & 1.8--2.06  & 1.88--2       & 2.9--3.06  & 1.88--2.06 & 2.89--3.01 \\
$k \ (\omega_{pe}/c)$    & 2--15	 & -15--0 & 0--2        & 5--15      & -10$\sim$-4   & 9--13      & -1.8--1.8  & -2.8--2.8 \\

\hline
\end{tabular}
}
\label{Tab:Table1}
\end{table}

\begin{table}[htbp]
\centering
\caption{The intensity ratios of $\rm H_{2}/H_{3}$ and $\rm H_{3}/H_{4}$ at saturation for Cases R and A--E. The energy ratios of the forward- (f) and backward- (b) propagating portions for $\rm H_2$, $\rm H_3$, and $\rm H_4$ are also listed.}
\setlength{\tabcolsep}{5mm}{
\begin{tabular}{ccccccccc}
\hline
\hline
                       & Case A       & Case B  & Case R & Case C & Case D & Case E   \\
\hline
$\rm W_{H2}/W_{H3}$     & 206.8 & 169.3  & 120.0      & 58.3     & 44.1       & 124.8   \\
$\rm W_{H3}/W_{H4}$     &  --	  &  --      &  --          & 93.6     & 88.5       &  --      \\
$\rm W_{H2f}/W_{H2b}$   & 2.8	  & 3.0    & 1.4        & 1.6      & 0.8        & 4.3       \\
$\rm W_{H3f}/W_{H3b}$   & 0.3	  & 0.7    & 4.0        & 7.5      & 8.5        & 1.1       \\
$\rm W_{H4f}/W_{H4b}$   &  --	  &  --      & --           & 1.4      & 3.7        &  --     \\

\hline
\end{tabular}
}
\label{Tab:Table2}
\end{table}

\end{CJK*}
\end{document}